# Interfacial chemistry meets magnetism: comparison of Co/Fe$_3$O$_4$ and Co/α-Fe$_2$O$_3$ epitaxial heterostructures.

*Ewa Madej* [*,1], *Natalia Kwiatek-Maroszek* [1,2], *Kinga Freindl* [1], *Józef Korecki* [1],
*Ewa Młyńczak* [1], *Dorota Wilgocka-Ślęzak* [1], *Marcin Zając* [2], *Jan Zawała* [1], *Nika Spiridis* [1]

[1] Jerzy Haber Institute of Catalysis and Surface Chemistry Polish Academy of Sciences, Niezapominajek 8, 30-239 Krakow, Poland

[2] National Synchrotron Radiation Centre SOLARIS, Jagiellonian University, Czerwone Maki 98, 30-392 Krakow, Poland

**Abstract**

The magnetic and chemical structure of metal/oxide interfaces were studied in cobalt/magnetite (Fe$_3$O$_4$) and cobalt/hematite (α-Fe$_2$O$_3$) epitaxial heterostructures using the comprehensive selection of microscopic and spectroscopic methods. It was observed that the cobalt nanostructures and ultrathin films were oxidized at both interfaces, with a thicker cobalt oxide layer in the system with hematite. The formation of cobalt oxides was accompanied by the interfacial reduction of iron that modified magnetic properties of the iron oxides layers. In particular, uncompensated magnetic moments appear in antiferromagnetic hematite, and the orbital magnetic moment of Co grown on magnetite is significantly enhanced for thicknesses below 1 nm. Synchrotron magnetic microscopy showed a direct correlation in the domain structures of the cobalt/iron oxides: ferromagnetic coupling between cobalt and magnetite and between cobalt and the magnetically modified layer of hematite.

## 1. Introduction

Metal-oxide interfaces play an important role in shaping the structural, electronic, and magnetic properties of functional heterostructures, which are widely studied due to their importance for both fundamental and applied sciences [1] in various fields, including spintronic [2–5] and catalysis [6,7]. Of special interest for spintronics are the heterostructures comprising magnetic oxides, both ferro- or ferrimagnetic (FM or FiM) and antiferromagnetic (AFM) [8]. In particular, FiM magnetite was one of the first oxides considered for spintronic applications due to its high Curie temperature and high spin polarization at the Fermi level [9]. Hematite has recently attracted much attention since its applicability in antiferromagnetic spintronic devices was demonstrated [10]. Numerous papers were devoted to metal-oxide systems, including an FM metal and a simple AFM oxide (see [11] for review). Still, much less attention was paid to the interfacial atomic and magnetic structure in epitaxial metal-oxide heterostructures, including the iron oxides mentioned above, i.e. $Fe_3O_4$ and $\alpha$-$Fe_2O_3$.

In this paper, we present a comparative analysis of the interfacial properties of two epitaxial systems: Co/$Fe_3O_4$(111) and Co/$\alpha$-$Fe_2O_3$(0001). It is worth noting that there is a lack of experimental data on the adsorption and growth of cobalt on the (111) surface of magnetite, and only a single paper deals with Co adsorption and magnetism of the Co/$Fe_3O_4$(001) system [12], as reviewed by Parkinson [13]. Conversely, cobalt films on α-$Fe_2O_3$(0001) films grown on Pt(111) and α-$Al_2O_3$(0001) single crystals have been extensively studied by a French group [14–19], who reported several fundamental structural and magnetic properties of this model FM/AFM thin film system.

Here, we present the significant refinement of that research by considering the so-called biphase surface structure of the studied oxides [20–23], introducing an alternative substrate for the iron oxide films, namely ultrathin Pt(111) films on MgO(111) [24], and examining the Co-thickness dependence. The comprehensive characterization of the Co-magnetite and Co-hematite systems grown by molecular beam epitaxy (MBE) was performed *in situ*, by surface sensitive methods, low energy electron diffraction (LEED) and scanning tunneling microscopy (STM), and *ex situ*, using methods with chemical and magnetic sensitivity: conversion electron Mössbauer spectroscopy (CEMS), X-ray absorption spectroscopy (XAS), X-ray photoemission electron microscopy (X-PEEM) and X-ray magnetic circular dichroism (XMCD). We demonstrated the direct correlation between the chemical and magnetic state of interfacial Fe and Co atoms.

## 2. Experimental details

The laboratory *in situ* experiments were performed in an ultrahigh vacuum (UHV) system with a base pressure of $2 \cdot 10^{-10}$ mbar, including MBE equipment and standard surface characterization tools, LEED, and STM. Among other metals, cobalt, and iron isotope ($^{57}$Fe) were evaporated from Knudsen cells (BeO crucibles, pressure during the deposition in a $10^{-10}$ mbar range) to grow metallic films or reactively, under molecular oxygen pressure, for oxide films. Platinum and MgO were evaporated from electron beam sources. The evaporation rate was calibrated using a quartz microbalance. Epitaxial iron oxide layers with a typical thickness of 10 – 20 nm were grown on two substrate types: on a Pt(111) single crystal and MgO(111) with a 10 nm thick epitaxial Pt(111) buffer layer. The single crystalline Pt(111) substrate was cleaned by repeated cycles of annealing in an oxygen atmosphere ($3·10^{-7}$ mbar, 10 min, 800 K), $Ar^+$ bombardment ($3·10^{-6}$ mbar, 1 kV, 10 mA, 30 min) and flashing at 1200 K under UHV until a sharp Pt(111)-(1x1) LEED pattern was observed. The Pt(111) films on MgO(111) were deposited at room temperature (RT) and annealed for 25 min at 800 K. Using the Pt(111)/MgO(111) substrates allowed a fast-track fabrication of good quality samples without the cumbersome process of cleaning the Pt single crystal and then simplified *ex situ* measurements. Iron oxide films were grown using the $^{57}$Fe isotope to enable Mössbauer spectroscopy characterization.

Epitaxial magnetite $Fe_3O_4$(111) films were grown by reactive deposition of iron on the Pt(111) surface under an oxygen partial pressure of $8·10^{-6}$ mbar at 520 K, followed by 30 min annealing at 770 K. Hematite, α-$Fe_2O_3$ (0001), films were prepared by oxidation of the pre-deposited magnetite films by annealing at 730 K for 45 min under an oxygen partial pressure of $3·10^{-5}$ mbar. At each preparation step, the film quality was checked *in situ* by LEED and STM.

Cobalt films, flat and wedged, were grown by MBE. The Co wedges were deposited at RT with a coverage gradient ranging from 0.2 to 2.2 nm over a 2-mm distance, concluded with a 3 nm flat film. The samples included reference areas without Co. All films were covered with a 3 nm MgO layer for *ex situ* measurements.

The STM measurements were performed at RT using an RHK VT-UHV300 microscope operating in the constant-current mode. The stoichiometry of the iron oxide films was verified by *ex situ* CEMS. CEMS measurements were performed at RT using a standard constant acceleration Mössbauer spectrometer, a He/$CH_4$ gas flow electron detector, and a 50 mCi $^{57}$Co/Rh source.

The synchrotron part of the research, X-PEEM and XAS experiments that allowed the elemental-sensitive chemical analysis and both exploiting the magnetic sensitivity of XMCD was performed at the soft X-ray bending magnet PEEM/XAS beamline [25] at the National Synchrotron Radiation Centre Solaris [26]. The beamline was equipped with two end stations: a PEEM station (Elmitec PEEM III microscope with the energy analyzer) and a universal XAS station.

PEEM imaging was performed at the Fe and Co $L_3$ edges with the left and right elliptically polarized X-rays illuminating the sample at a grazing angle of 16º. The magnetic contrast of FM and FiM domains emerges as the difference, pixel by pixel, of two PEEM images, $I_-$ and $I_+$, taken with the opposite helicities, normalized to their sum, to yield the XMCD-PEEM asymmetry image $I_{XMCD} = \frac{I_- - I_+}{I_- + I_+}$. The local XMCD asymmetry is proportional to the projection of the magnetization on the direction of the incident X-ray beam.

In the XAS station, the XMCD spectra were derived from the XAS measurements at the Fe and Co $L_3$ edges in the external magnetic field of ±0.14 T parallel to the X-ray beam, and the photon incident angle was 45º for left and right elliptical polarization. The degree of polarization was determined using thick Fe and Co samples and literature data concerning experimental verification of the XMCD sum rule for iron and cobalt [27]. In the XAS station, the spectra could be measured with a spatial resolution of 40 μm, whereas the PEEM images were collected down to the 10 μm field of view (FoV), with a theoretical spatial resolution of one pixel out of a 512 x 512 array.

3. Results and discussion

3.1. Characterization of the oxide layers

The surface structure, represented by LEED patterns, and morphology, observed from STM images, of the two substrate types used for the preparation of the iron oxide layers are shown for the Pt(111) single crystal and the Pt(111)/MgO(111) substrates in Figs 1a and 1e, respectively. The surface of the single crystal is characterized by large, flat terraces with uniformly oriented monoatomic step edges and a sharp LEED pattern with 3-fold symmetry of the (111)-fcc surface. The average terrace width of approximately 15 nm indicates a miscut angle between 0.5 and 1 degree relative to the (111) plane. The STM morphology of the 10 nm Pt layer on MgO(111), shown in Fig. 1e, is different. The less regular terraces are separated by a pair of coupled screw dislocations. This morphology results in blurred LEED spots, reflecting the distribution of local terrace inclination over the probed surface.

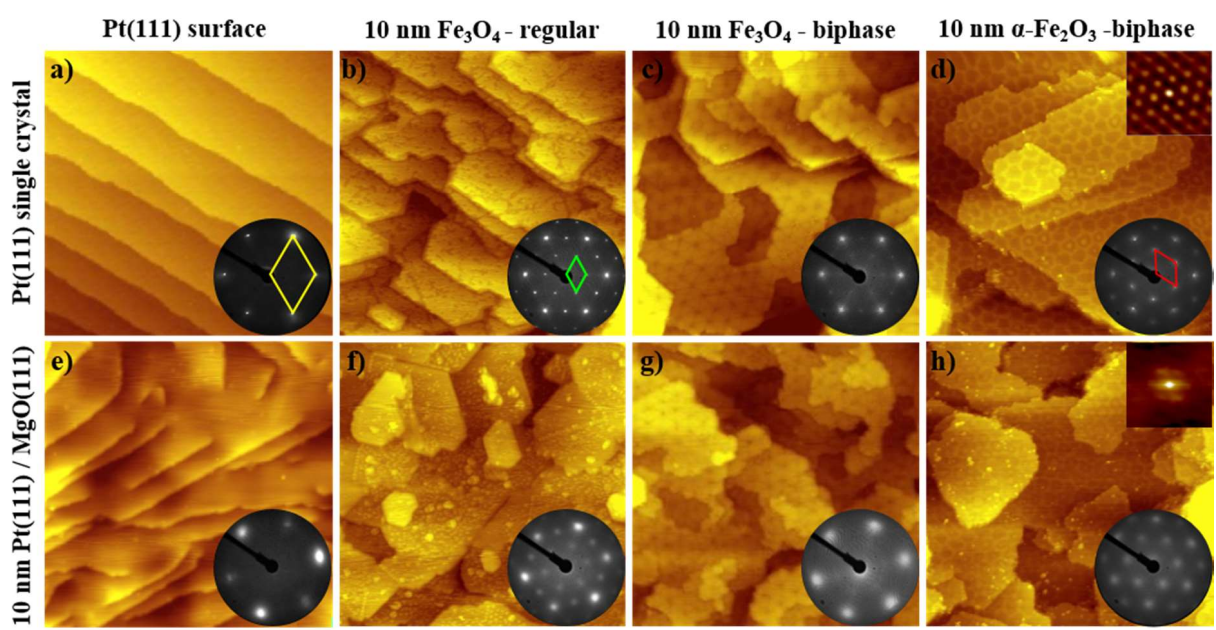

Fig. 1 STM images (85 x 85 nm$^2$) of the iron oxide layers prepared on Pt(111) single crystal (top row) and the Pt(111)/MgO(111) substrates (bottom row), bare substrates (1$^{st}$ column), 10 nm Fe$_3$O$_4$ (2$^{nd}$ column), biphase on Fe$_3$O$_4$ (3$^{rd}$ column), and 10 nm α-Fe$_2$O$_3$(0001) (4$^{th}$ column). For each STM image, the corresponding LEED pattern taken at an electron energy of 90 eV, with the reciprocal lattice unit cells marked, is shown in the inset, as well as autocorrelation patterns (25 x 25 nm$^2$) from STM images of the hematite surfaces.

The morphological differences between the Pt-substrates are reflected in the corresponding images of the magnetite layers, as shown in Fig. 1b and 1f for 10 nm $Fe_3O_4$ deposited on the Pt(111) single crystal and Pt(111)/MgO(111), respectively. In both cases, the STM images reveal continuous films with a step height corresponding to the physical monolayer thickness of $Fe_3O_4$(111), i.e. 0.5 nm. The LEED patterns exhibit a (1x1) hexagonal symmetry of the so-called regular $Fe_3O_4$(111) surface [20]. The stoichiometric magnetite films with this regular surface were the starting point for stabilizing the $Fe_3O_4$(111) surfaces with biphase superstructures or obtaining hematite films. According to our standard procedure [20], the biphase superstructure was stabilized by the surface enrichment of the magnetite film with 0.4 nm of metallic Fe deposited at RT, followed by annealing at 720 K for 15 minutes. For magnetite deposited on the Pt(111) single crystal (Fig. 1c), the biphase appears in the LEED pattern as satellites surrounding the main magnetite spots. In the corresponding STM image, two types of hexagonal superstructure are seen on different terraces, with a periodicity of 5 ± 0.5 nm, which means a coexistence of two biphase superstructures termed in the literature as B and C [21]. In contrast, the blurred spots for the Pt(111)/MgO(111) substrate hinder observation of the biphase superstructure in the LEED pattern. However, a hexagonal network similar to that observed on the Pt(111) single-crystal substrate is distinct in the STM image shown in Fig. 1g.

Results for hematite layers formed by the oxidation of the magnetite films are shown in Fig. 1d and 1h. The LEED patterns follow the tendency observed for the magnetite films. For the single crystal Pt(111) substrate (Fig. 1d), sharp spots with satellites of the biphase superstructure [22,28] indicate the high structural quality of the α-$Fe_2O_3$(0001) surface, and for the Pt(111)/MgO(111) substrate (Fig. 1h) the pattern symmetry unambiguously indicates the α-$Fe_2O_3$(0001) surface but the spot broadening leaves the structural details hidden. In turn, the STM images of the hematite layer on Pt(111) (Fig. 1d) and Pt(111)/MgO(111) (Fig. 1h) both show surfaces with a honeycomb biphase superstructure. For hematite on Pt(111), the superstructure periodicity is 4.5 nm, whereas, for hematite on Pt/MgO(111), an irregular superstructure has periodicity from 4.0 nm up to as high as 6.5 nm. In contrast to the magnetite films, the uniformity of the biphase in the hematite films on the Pt single crystal is distinctly superior to that on the Pt buffer layer on MgO(111).

Characterization of the oxide films beyond the surface layer probed by LEED and STM was accomplished using CEMS, which probes the entire film volume. The results of the CEMS measurements are summarized in Fig. 2. The room temperature CEMS spectrum of a 10 nm $Fe_3O_4$(111) layer on Pt(111)/MgO(111) in Fig. 2 (top) indicates a hyperfine pattern of bulk magnetite: two six-line magnetic components, red and green, that correspond to the iron ions in tetrahedral ($Fe^{3+}$) and octahedral ($Fe^{2+}$, $Fe^{3+}$, averaged due to electron hopping to $Fe^{2.5+}$) sites, respectively, are occupied in the 1:2 ratio corresponding to the perfect stoichiometry of magnetite [29]. The relative intensity of the sextet lines depends on the angle $\theta$ between the hyperfine magnetic field (local magnetization) and the propagation direction of the γ-rays, and is given by 3:x:1:1:x:3, where $x = \frac{4sin^2\theta}{2-sin^2\theta}$. For example, for the given geometry of the CEMS measurements (γ-rays along the sample normal) the in-plane and perpendicular magnetization would result in x = 4 and x = 0, respectively. The experimental value $x = 3.4 \pm 0.1$ perfectly fits the homogenous magnetization distribution over eight <111> magnetization easy directions for magnetite, six of which are at the angle of 70.5 ° from the sample normal and two along the normal. Apparently, the magnetocrystalline anisotropy dominates over the shape anisotropy that would favor the in-plane magnetization. Low-temperature CEMS measurements (not shown here) indicated that the Verwey transition for the 10 nm magnetite film is at $124 \pm 2$ K, in agreement with earlier results [28].

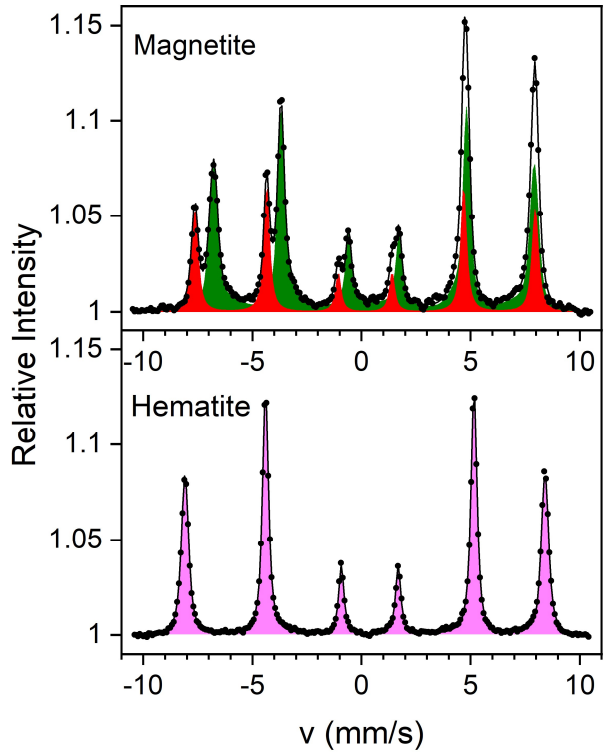

Fig. 2 The CEMS spectrum of $Fe_3O_4(111)$ on $Pt(111)/MgO(111)$ (top) and $\alpha\text{-}Fe_2O_3(0001)$ on $Pt(111)/MgO(111)$ (bottom). Both oxide films are approximately 10 nm thick.

The CEMS spectrum of 10 nm $\alpha\text{-}Fe_2O_3(0001)$ film (Fig. 2, bottom) reveals perfect $Fe^{3+}$ sites of bulk hematite. The relative intensity parameter $x = 4$ unambiguously indicates an in-plane orientation of the AFM spins. The hematite film did not exhibit the Morin transition down to 100 K, the lowest achievable measurement temperature. This observation is not surprising considering the sensitivity of the Morin transition to epitaxial strains [30].

3.2. Chemical and magnetic properties of cobalt films and cobalt-iron oxide interfaces

3.2.1. Comparative STM analysis of Co on magnetite and hematite

Magnetite and hematite surfaces shown in Fig 1c, d, and h were the templates for the deposition of cobalt. The topographic STM images as a function of the Co coverage deposited at RT on $Fe_3O_4(111)/Pt(111)$, $\alpha\text{-}Fe_2O_3(0001)/Pt(111)$, and $\alpha\text{-}Fe_2O_3(0001)/Pt(111)/MgO(111)$

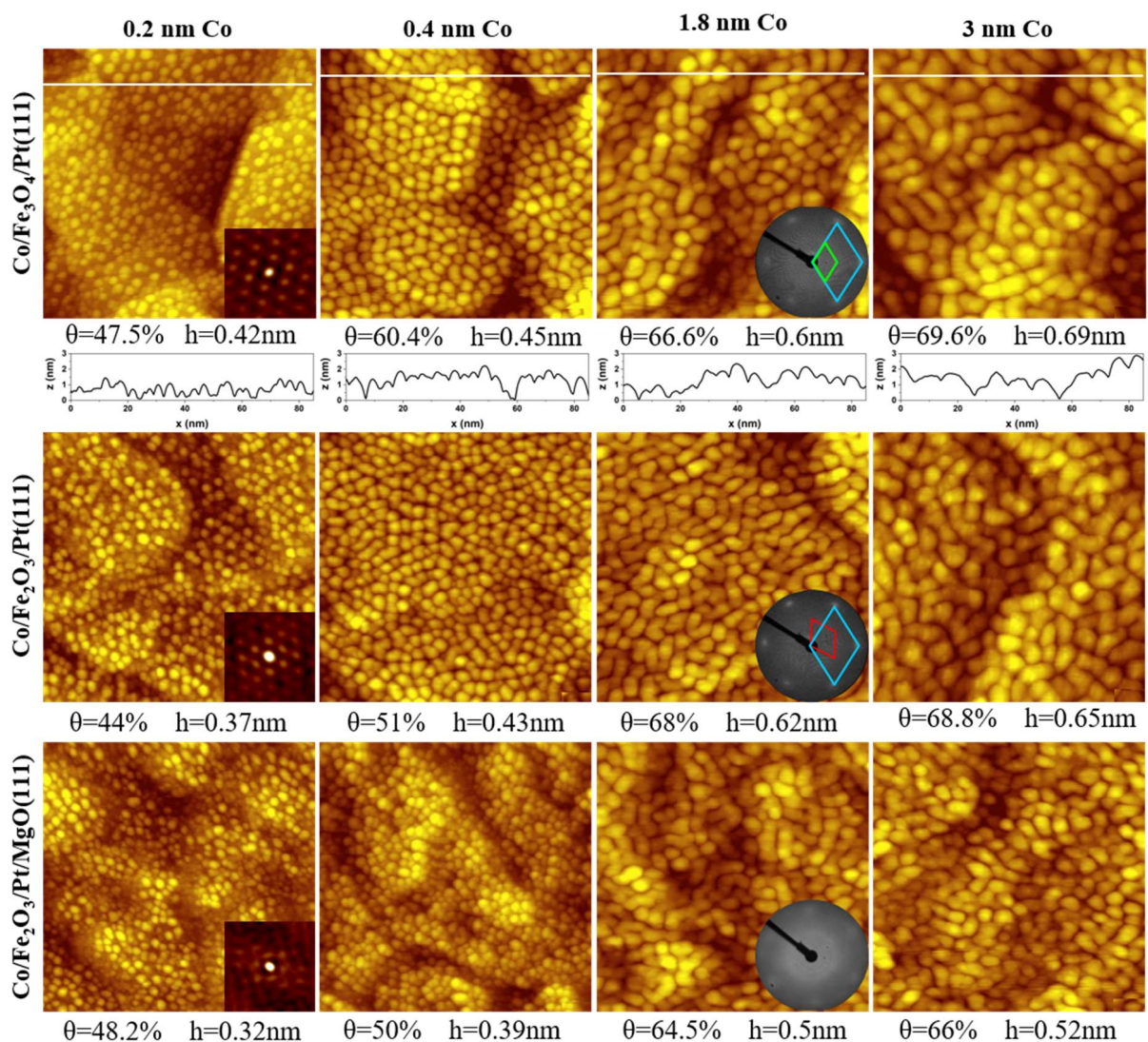

Fig. 3. STM images (85 x 85 nm$^2$) of the Fe$_3$O$_4$(111)/Pt(111) (top row), α-Fe$_2$O$_3$(0001)/Pt(111) (middle row), and α-Fe$_2$O$_3$(0001)/Pt(111)/MgO(111) (bottom row) surfaces covered by increasing amount of Co deposited at RT. The cobalt surface coverage and average height of the nanostructures are shown below each STM image. The insets show an autocorrelation pattern (25 x 25 nm$^2$) from STM images of 0.2 nm Co and LEED pattern (90 eV) with marked cobalt and iron oxides unit cells.

are collected in Fig. 3 in the top, middle and bottom row, respectively. Taking into account the similarity between Fe$_3$O$_4$(111)/Pt(111) and Fe$_3$O$_4$(111)/Pt(111)/MgO(111) surfaces (compare Figs 1c and 1g), STM imaging of cobalt on this second surface was abandoned. The uniform distribution of the cobalt clusters over all surfaces proves homogenous nucleation. However, for the lowest cobalt coverages (0.2 nm), the role of the biphase superstructure for nucleation becomes apparent in some areas, where the Co clusters are arranged in a hexagonal pattern corresponding to the biphase periodicity. This is confirmed by autocorrelation patterns taken from the STM images for 0.2 nm Co and shown in the insets of Fig. 3. The autocorrelation patterns indicate the presence of short and long-range order by displaying the distances between repeated features found in the STM images. The periodicities of cobalt nanoparticles on $Fe_3O_4(111)/Pt(111)$ and $\alpha$-$Fe_2O_3(0001)/Pt(111)$ determined based on the autocorrelation patterns were 5.1 nm and 4.4 nm, respectively. These values roughly agree with the biphase's periodicity on both iron oxides, as determined from our STM images.

The phase of homogeneous nucleation at characteristic structural sites of the biphase ends for both substrates already at the lowest coverages (0.2 nm). This initial phase is followed by the growth of three-dimensional islands, which subsequently undergo coalescence. As the thickness of cobalt deposits increases, the hexagonal LEED patterns observed on the Co-covered surfaces (illustrated in the insets of Fig. 3 for 1.8 nm Co) show the reciprocal space surface unit cells (2x2) or ($\sqrt{3}\times\sqrt{3}$) R30° (marked in blue) relative to the underlying magnetite or hematite cells (marked in green or red), respectively. Due to the significant lattice mismatch between cobalt and both magnetite (15.5%) and hematite (13.7%), the expectation for pseudomorphic epitaxy is unreasonable. Most probably, the adjustment of the (111)-oriented Co layers to the oxide substrates occurs through dislocation networks, as proposed earlier for the Co/$\alpha$-$Fe_2O_3(0001)$/Pt(111) system [17]. The sample cross-sections shown below the STM images for Co/$Fe_3O_4$/Pt(111) illustrate the change in the Co adsorbate morphology with increasing coverage. A quantitative analysis of island dimensions and their statistics was performed using the automated segmentation algorithm in the SPIP$^{TM}$ (Scanning Probe Image Processor) STM software package developed by Image Metrology. The significant data of this analysis, i.e. the average height of the Co islands and their surface coverage, are given below the corresponding figures in Fig. 3. The island heights, as determined from the STM images compared to the nominal amount of deposited cobalt, indicate a transition to the three-dimensional nanoparticle growth only up to a thickness of 0.4 nm. Beyond this threshold, growth shifts to the formation of islands atop a quasi-continuous cobalt (q-Co) layer. Figure 4 illustrates the morphology of the Co adsorbate by comparing the nominal cobalt deposition amount (represented by a dashed line) with the amount of cobalt contained in the islands (full symbols) and within the q-Co layer (indicated by open symbols). First, it can be noted that the growth behavior of Co on both iron oxides is very similar. The equivalent thickness of material present in the islands is proportional to the product of the coverage and the average height of the islands, whereas the thickness of q-Co is the result of subtracting the layered equivalent of the islands from the nominal Co thickness.

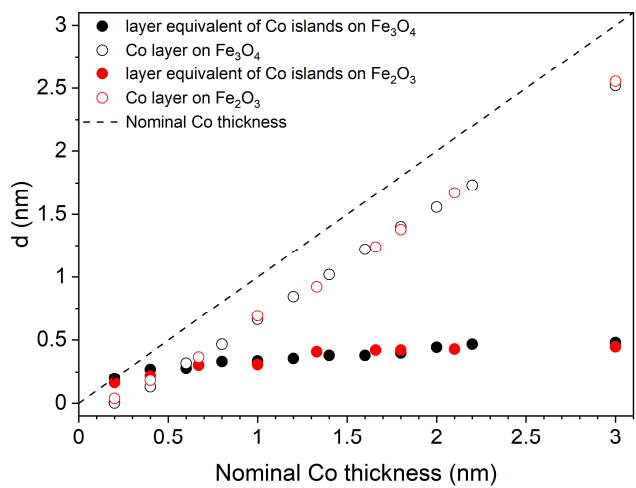

Fig. 4 Nominal thickness of Co deposit (dashed line), layer equivalent of cobalt contained in islands (full symbols), and thickness of quasi-continuous Co layer (open symbols) on magnetite (black) and hematite (red).

Analysis of Figures 3 and 4 reveals that the islands nucleate as bilayer objects, based on the reasonable assumption that the Co layer thickness is approximately 0.2 nm. However, the STM images do not provide insights into the chemical nature of the Co-oxide interface. This aspect will be discussed below based on synchrotron measurements, which, as demonstrated by Bezencenet et al. [15] for the Co/hematite system, can effectively detect oxidation states even in covered Co layers. Summarizing, the STM analysis indicates that the type of oxide substrate does not significantly influence the morphology of the cobalt layers. In both studied cases, the initial growth of islands and their subsequent coalescence result in forming a characteristic structure of quasi-continuous Co film.

3.2.2. Chemical structure at Co/magnetite and Co/hematite interfaces - comparative XAS analysis

The chemical structure at the cobalt/iron oxides interfaces was studied using XAS measurements on the samples with a gradient of the Co thickness. The shape of the XAS spectra and the peak position depend on the local electronic structure, and from the characteristic spectral features, it is possible to distinguish between the metal and oxide states of a given element. Figures 5a and 5b show the Co $L_3$ edge spectra measured for increasing cobalt thicknesses deposited on magnetite and hematite, respectively. For the thinnest films, up to 0.5 nm Co, the XAS spectra reveal multiple features characteristic of oxidized Co [31]. Then, the spectra gradually evolve towards a single $L_3$ peak, characteristic of Co metal. Deconvolution of the spectra measured for 0.2 nm Co into the spectrum of pure Co [32] and CoO [33] (Fig. 5a) showed that approximately 60% of the 0.2 nm Co film is oxidized on the magnetite surface. In contrast, for hematite, the oxidized portion of the corresponding Co film amounts to 90% (Fig. 5b). The fast decrease of the Co-oxide signal with increasing cobalt thickness shows that the oxidation is likely to be limited only to the interface region. It should be mentioned that the XAS spectrum of cobalt spinel (CoFe$_2$O$_4$) [34] is practically indistinguishable from CoO, and incorporation of Co in the magnetite structure, albeit less probable at RT, cannot be excluded.

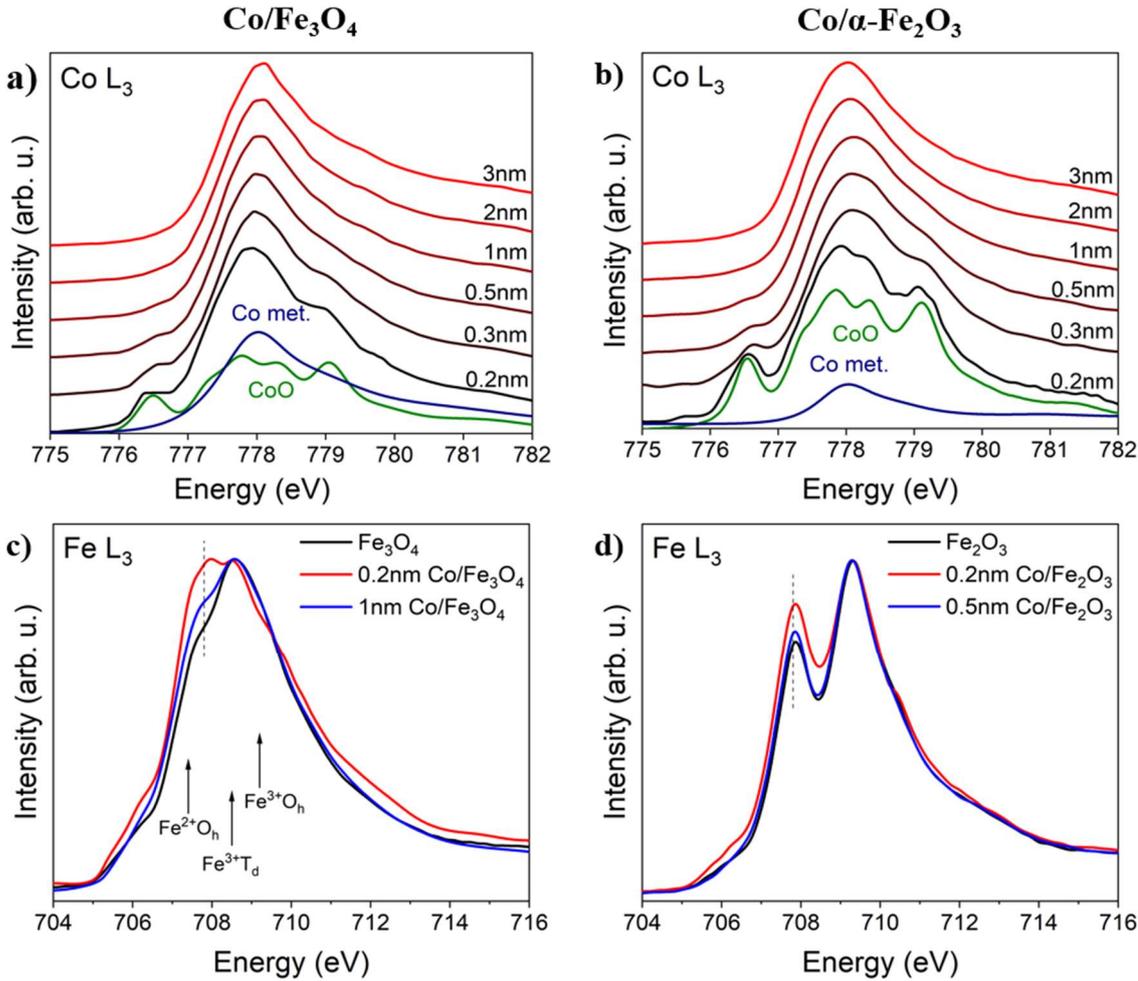

Fig. 5. Co L$_3$ edge (a, b) and Fe L$_3$ edge (c, d) XAS spectra measured for the increasing amount of cobalt deposited on magnetite and hematite substrates, (a,c) and (b,d), respectively. For the 0.2 nm Co L$_3$ spectra, deconvolution into metal-Co and CoO is shown by the blue and green lines, respectively.

As we show by measuring XAS spectra on the Fe L$_3$ edge, cobalt oxidation is accompanied by interfacial reduction of iron oxides. Figures 5c and d compare XAS spectra for bare magnetite and hematite (black curves) with those with the Co overlayers. For the thinnest Co coverage (0.2 nm, red curves), when the sensitivity to the interfacial oxide layer is the highest, an increase in the intensity close to 708 eV is observed for both oxides. This energy is associated with iron in the second oxidation state [35]. For higher Co coverages (Fig. 5c, d blue curves), the XAS spectra are approaching the spectra of pure magnetite and hematite.

Summarizing, the XAS measurements proved the interfacial oxidation of the Co layer accompanied by the interfacial reduction of iron oxides. The resulting interfacial oxide phases should not be interpreted as bulk compounds but as local deviation from perfect stoichiometry caused by the deficiency of oxygen atoms. Such a picture is supported by our CEMS measurements (for details, see Supplementary materials SM 1), which do not show any distinct differences in spectra measured for iron oxide without and with the Co over-layers. The effect of Co is seen only as a minor broadening of the characteristic hyperfine patterns, indicating the presence of Fe species in a lower oxidation state.

3.2.3. Magnetic structure of Co/magnetite and Co/hematite heterostructures imaged using XMCD-PEEM

The chemical processes at the interfaces strongly influenced the magnetic properties of the metal-oxide systems under investigation, as shown using PEEM imaging with the synchrotron X-ray excitation. This was possible because the X-PEEM technique enables the direct observation of the FM or FiM domain structures with elemental sensitivity by matching the photon energy to the absorption edge of the given element and optimizing the magnetic contrast by fine tuning the energy to characteristic features of the XAS spectra for magnetite [36,37] and hematite [16].

3.2.3.1 Co/magnetite

Figure 6 shows XMCD-PEEM images of the magnetic domain structure recorded at the same area, at the Co $L_3$ edge (top row) and Fe $L_3$ edge (bottom row) for increasing thickness of Co (0.5, 1, and 2 nm) deposited on 10 nm Fe$_3$O$_4$ on Pt(111)/MgO(111). The domain structures of magnetite and cobalt are precisely the same, indicating the ferromagnetic coupling between Co and Fe$_3$O$_4$.

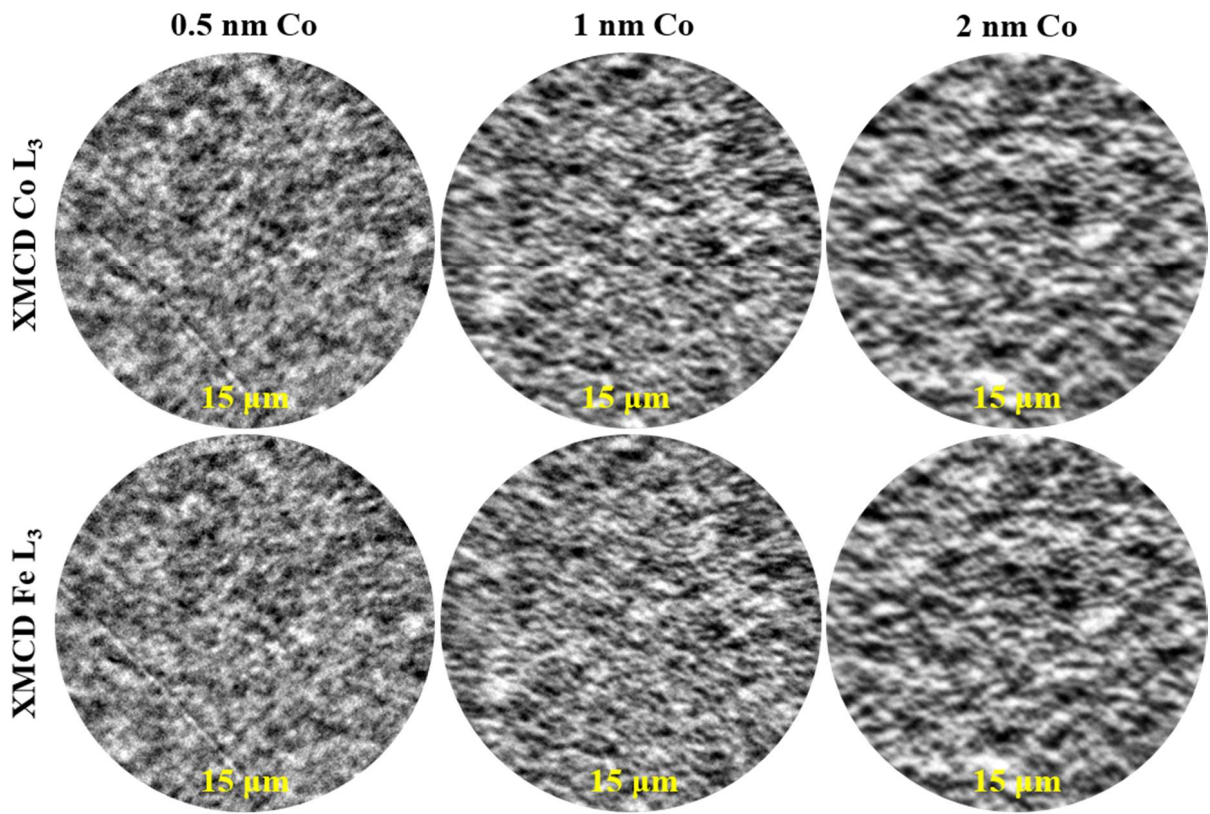

Fig. 6. XMCD-PEEM images (FoV 15 μm) recorded at the Co L$_3$ edge (top row) and Fe L$_3$ edge (bottom row) for different amounts of Co (0.5, 1, and 2 nm) deposited on 10 nm Fe$_3$O$_4$/Pt(111)/MgO(111). Photon energy was 708.2 eV and 778.0 eV for the Fe and Co L$_3$ edges, respectively.

The domains for magnetite without cobalt (see SM 2) are sub-micrometer-sized, and their character remains unchanged as the Co thickness increases in measured thickness range. A similar type of domain structure was observed before for ultrathin magnetite films using magnetic force microscopy [38,39], XMCD-PEEM [40], and it remains in strong contrast to the magnetic domains observed on the (111) surface of bulk magnetite [41]. The origin of the fine domains is attributed to a spin disorder at the anti-phase boundaries formed during the nucleation of the magnetite films [42]. In correlation with our CEMS results, the observed differentiated contrast distribution is interpreted as resulting from domains with in-plane and out-of-plane magnetization components. The in-plane magnetization components may exhibit six orientations along the <211> directions in the (111) plane, corresponding to the surface projections of the off-normal <111> magnetization easy axes.

As pointed out above, the domain structures of FiM magnetite and FM cobalt are the same. Their character did not change relative to pure magnetite, indicating that magnetite enforces the domain structure in the cobalt layer. Consequently, the size and shape of the cobalt magnetic domains do not change with the Co thickness in the investigated range. Moreover, the exchange interaction with magnetite stabilizes superparamagnetism that is expected for cobalt nanostructures [43]. This interaction makes the magnetic contrast visible even in the limit of the smallest cobalt deposit with a nominal thickness of 0.2 nm, which gives nanoparticles 2.2 nm in diameter and 0.42 nm in height (see SM 2).

3.2.3.2 Co/hematite

Figure 7 shows XMCD-PEEM images of the magnetic domain structure for increasing thickness of cobalt deposited on α-$Fe_2O_3$(0001)/Pt(111)/MgO(111). The magnetic domains in cobalt on hematite are significantly larger than those of cobalt on magnetite (compare Fig. 6). In contrast to cobalt on magnetite, the domain structure could not be detected below a Co thickness of 0.5 nm. This we interpret as a combined effect of superparamagnetism (suppressed for Co/magnetite) and the higher degree of oxidation (compare Fig. 5). Approximately 90 % of the 0.2 nm Co film is oxidized on the hematite surface, which explains why the magnetic domains are not observed below 0.5 nm of Co. The character of the domains did not depend significantly on Co thickness.

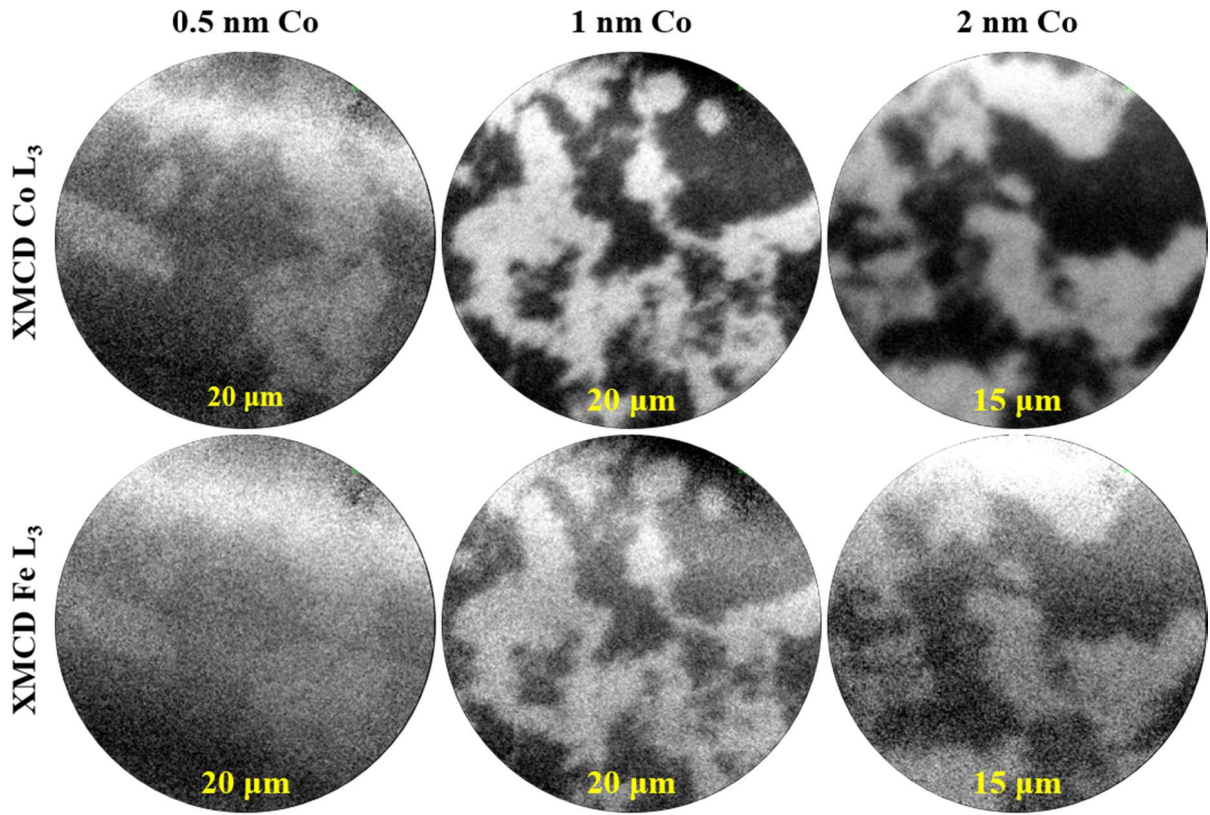

Fig. 7. XMCD-PEEM images recorded at the Co $L_3$ edge (top row) and Fe $L_3$ edge (bottom row) for different amounts of Co (0.5, 1, and 2 nm) deposited on 10 nm α-$Fe_2O_3$/Pt(111)/ MgO(111). Photon energy was 707.8 eV and 778 eV for the Fe $L_3$ and Co $L_3$ edges, respectively.

Most importantly, the XMCD contrast was also observed at the Fe $L_3$ edge (Fig. 7, bottom row), with the domain pattern identical to the Co domains. The XMCD contrast for nominally AFM hematite indicates the presence of uncompensated interfacial Fe spins. Bezencenet et al. [15], based on XAS/XMCD spectra, reported reduced hematite, contributing to an XMCD response, and oxidized cobalt at the Co/α-$Fe_2O_3$ interface. Now, we directly visualized the uncompensated magnetic moments on iron atoms that are ferromagnetically coupled with the cobalt atoms. Interestingly, the interfacial layer of cobalt oxide does not prevent direct ferromagnetic coupling between cobalt and the reduced iron oxide layer. This can be explained either by magnetic polarization of the cobalt oxide layer or, more plausible, by the discontinuous character of the Co oxide layer. Moreover, based on the XMCD-PEEM and X-ray magnetic linear dichroism PEEM observation of the same system [18], we can indirectly conclude that ferromagnetic spins and the corresponding domain structure correlate not only with the interfacial uncompensated Fe spins but also with the underlying AFM domains.

XMCD-PEEM imaging enables directional analysis of the magnetization that is based on the dependence of the local intensity of the XMCD image, $I_{XMCD}$, on the angle α between k-vector of the X-ray and the magnetic moments M: $I_{XMCD} = \langle M \rangle \cos(\alpha)$. Using this dependence, the in-plane distribution of the magnetization can be extracted from the image series collected by rotating the sample around the normal. The rotation leads to distinct contrast intensity changes for the domains with an in-plane magnetization component, whereas domains with the out-of-plane magnetization do not change their contrast. Due to the hexagonal symmetry, depending on the angle α, the cobalt domains should show at least three and a maximum six contrast levels. Dark and bright domains dominate the observed domain structure in Fig. 7. This can be roughly interpreted as a contrast grouping for domains with magnetization components parallel and antiparallel to the k-vector of X-rays. Consequently, the situation drastically changes upon rotation. The most characteristic images selected from the set collected by rotating the sample by 90° in eight steps are shown in Fig. 8. The contrast is reversed in some regions. Still, it remains almost unchanged in others - moreover, the degree of the contrast fragmentation changes toward smaller domains, whose boundaries become less apparent. To reconstruct a complete magnetization map, the expected angular dependence of the XMCD signal intensity was fitted for each pixel to determine the corresponding magnetization angle. Figure 8e shows a two-dimensional magnetization map of the ferromagnetic domain structure constructed from the angle-dependent XMCD-PEEM images. The histogram of magnetization azimuthal angles is shown in the polar graph in Fig. 8f. Indeed, the magnetization distribution exhibits features mostly concentrated in two opposite directions and substantial contribution around 60° of the main axis. Apparently, one of three easy axes, expected due to hexagonal symmetry, is not represented. The absence of the third easy axis may be attributed to the limited field of view of the images.

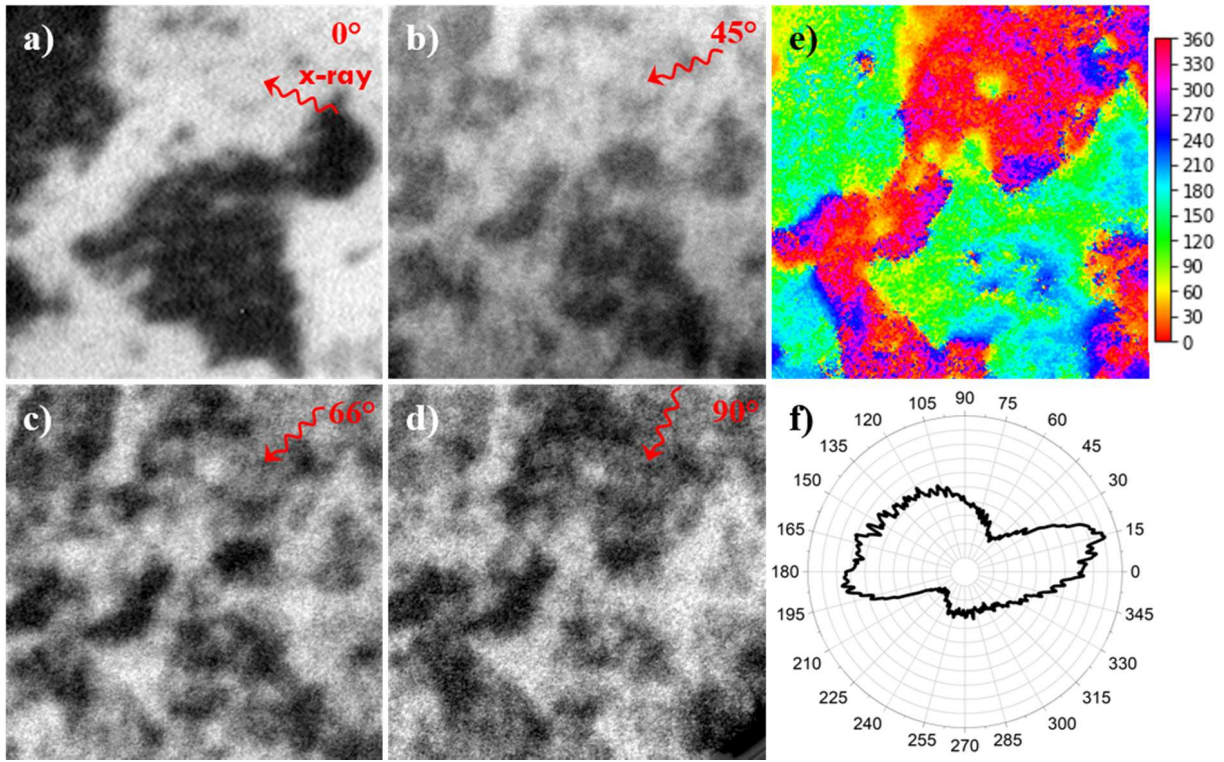

Fig. 8. XMCD-PEEM images (FoV 10 μm) recorded at the Co $L_3$ edge for 1 nm Co deposited on 10 nm α-$Fe_2O_3$/Pt(111)/MgO(111) at azimuthal angles of (a) 0°, (b) 45°, (c) 66°, (d) 90° between the incident synchrotron radiation beam direction and the [110] direction of the MgO substrate. The two-dimensional magnetization map (e) is constructed from angle-dependent XMCD-PEEM images. (f) Histogram showing the number of pixels in the image (e) characterized by a given azimuthal magnetization angle.

Concluding magnetic information from XMCD-PEEM, we observed that the domain structure of the ferromagnet (Co) is driven via direct exchange coupling by the anisotropy of the ferrimagnet ($Fe_3O_4$). In contrast, for the FM/AFM (Co/α-$Fe_2O_3$) system, it is rather that cobalt dominates over hematite, similar to another FM/AFM system, namely CoO/Fe(110) [33].

3.2.4. Comparative analysis of the magnetic moments in Co/magnetite and Co/hematite heterostructures using XMCD-XAS

To deepen the understanding of the magnetic structure at the interface between Co and the iron oxides, PEEM imaging of the domain structure was complemented by spectroscopic information from XMCD-XAS measurements. XMCD spectroscopy in core-level absorption, combined with the sum rules, allows an element-specific determination of orbital and spin magnetic moments [27], thus providing a powerful tool to study the magnetism of multi-component systems. The XMCD spectra at Co and Fe $L_{2,3}$ edges as a function of the thickness of the Co layer deposited on the magnetite and hematite films are shown in Fig. 9a and b, respectively. The intensity of the XMCD signal increases with the decreasing contribution of the CoO signal to the XAS spectra (compare Fig. 5a, b). It should be noted that both in PEEM and XAS, estimation of the investigated cobalt thickness is subjected to uncertainty that makes direct comparison somewhat ambiguous. With this in mind, the appearance of the XMCD signal correlates well with the minimum Co thickness for which the domain structure is observed: 0.2 nm for magnetite and 0.5 nm for hematite.

The increasing amplitude of the normalized XMCD Co signal reflects both the chemical changes of the interfacial cobalt layer and possible changes in the Co magnetic moments as a function of thickness. For both substrates, the XMCD amplitude saturates at approximately 1.5 nm Co, below which the XMCD signal is affected by chemical and magnetic size effects. Because the cobalt atoms in interfacial CoO do not contribute to the XMCD signal, the magnetic moments obtained from the sum rule analysis were normalized taking into account the relative contribution of the metallic Co for different thicknesses.

Figures 9c and d show the results of the sum rule analysis of the spin and orbital moments dependence of Co in the metallic phase as a function of thickness. The results for the magnetite and hematite substrates are essentially different. For magnetite (Fig. 9c) $m_{spin}$ exhibits only a weak thickness dependence: after initial growth from 1.45±0.06 $\mu_B$ to the maximum value of 1.79±0.05 $\mu_B$ for 1 nm Co, it moderately decreases and stabilizes at 1.59±0.04 $\mu_B$ for the thickest 3 nm Co deposit. The FM order is established for the thinnest films. However, the magnetic moments are still affected by superparamagnetism, which is responsible for the initial non-monotonous thickness dependence of the spin magnetic moment. Considering the morphology of the Co films, as observed by STM, it is highly probable that the magnetic field during the XMCD-XAS measurements is insufficient to saturate the smallest Co particles, which causes an underestimation of the magnetic moments. This effect is much less pronounced for the orbital magnetic moment that starts at 0.47±0.06 $\mu_B$ and monotonously decreases to 0.17±0.02 $\mu_B$. The pronounced enhancement of the orbital moment in the thinnest film limit is even better exposed in the $m_{ratio}= m_{orb}/m_{spin}$ that reaches a value as high as $m_{ratio} = 0.35±0.02$ for the thinnest film and then monotonously decreases to 0.1±0.01, as shown Fig. 9c. This analysis indicates that the size effects are limited to the Co film below 1.5 nm, and the magnetic moments for our thickest 3 nm film are close to the values of the bulk material [27]. A similar conclusion can be drawn for cobalt on hematite, although due to the lack of FM coupling to the substrate, the superparamagnetic effects extend to thicker layers. Therefore, reliable sum rule analysis is possible starting from a thickness of 0.5 nm. Consequently, the enhancement of the orbital magnetic moment is less pronounced, but the size effect vanishes for similar cobalt thicknesses. The size effect of the orbital magnetic moment enhancement for ultrathin cobalt films and multilayers was predicted theoretically [44,45] and observed experimentally using XMCD measurements [44,46–50]. Most of the cited papers are concerned with cobalt interfaced with a nonmagnetic metal, especially with Cu, for which theory and experiment agree about several percent enhancement of mspin and much stronger, up to 100 % enhancement of morb in the monolayer limit of the Co thickness. This size effect rapidly decreases with thickness d as 1/d, which reflects its surface/interface character [44].

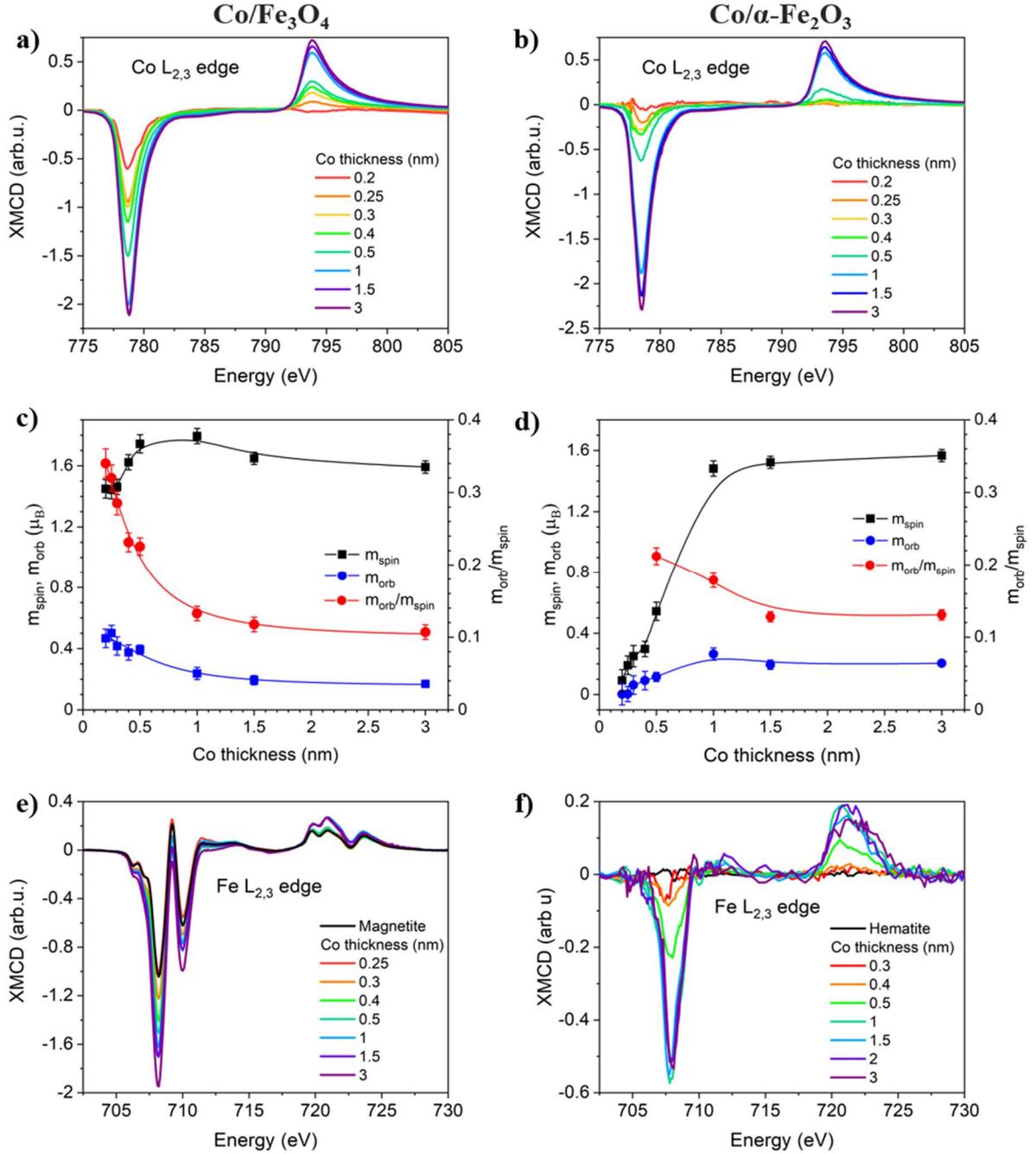

Fig. 9. (a) and (b) Co $L_3$ XMCD spectra for increasing cobalt thickness on magnetite and hematite, respectively, and the corresponding results of the sum rule analysis, (c) and (d), respectively. The lines in (c) and (d) are guides to the eye. (e) and (f) Fe $L_3$ XMCD spectra for increasing cobalt thickness on magnetite and hematite, respectively. For the Fe edge, XMCD spectra for the clean $\alpha$-$Fe_2O_3$ and $Fe_3O_4$ substrates are also shown.

Recently, Zhang et al. [48] reported XMCD experiment and related first principle calculation for Co/oxide systems, namely Co/Fe$_3$O$_4$(001) and Co/MgO(001), the former being closely related to our Co/Fe$_3$O$_4$(111). They observe a surprisingly high enhancement of m$_{orb}$ for a Co film as thick as 3 nm, which was our upper limit, and for which the size effect practically disappeared in the presently reported experiment. Without going into details about the significant discrepancy in the extent of the size effects between the data in Zhang's et al. work and the present results, it seems that the enhancement of the orbital magnetic moment in the Co/iron-oxide system is significantly larger than that of Co/metal. The source of this effect should be sought in the specific electronic structure of the Co/iron-oxide interface.

Figure 9e shows XMCD spectra for the magnetite substrate at the Fe L-edge as a function of the Co thickness, including also the spectrum of Co-uncoated film (black curve). All spectra present typical bulk Fe$_3$O$_4$ features of the L$_3$ edge, i.e. three lines at the energies 708.3 eV, 709.4 eV, and 710.1 eV, which are related to Fe$^{2+}$(O$_h$), Fe$^{3+}$(T$_d$) and Fe$^{3+}$(O$_h$), respectively [36]. Such a complex structure is associated with antiparallel spin orientations at octahedral and tetrahedral sites. With increasing Co thickness, we observed two effects: (i) a pronounced increase in the XMCD signal and (ii) a change in the relative intensities of the three-line structure. We interpret the intensity increase as the effect of easier magnetic saturation of magnetite beneath cobalt due to ferromagnetic coupling between easy-saturating cobalt and magnetite. This observation is supported by Kerr magnetometry (see SM 3), which shows that magnetization of the magnetite film without Co is not saturated in the magnetic field as high as 1 T, whereas the experimental setup of the Solaris XAS station gives the possibility to apply maximum magnetic field of only 0.2 T. On the other hand, there is a striking difference between XMCD spectra measured in field and in remanence for magnetite with and without Co. The second effect of reducing the line intensity at 709.4 eV, associated with the $Fe^{3+}$ contribution, can be quantitatively explained as a surface reduction of magnetite by cobalt.

Fig. 9f shows XMCD spectra of Co/α-$Fe_2O_3$ at the Fe edge, including the spectrum of Co-uncoated film (black curve). For clean hematite, there is no XMCD signal. On the other hand, in agreement with the magnetic domain structure observed at the Fe L-edge (compare Fig. 6), a relatively strong XMCD signal appears for increasing Co thickness. This signal and the observed domain structure reflect an FM or FiM Fe-composed layer at the Co-hematite interface. Bezencenet et al. [15,16] interpreted a similar XMCD signal for the same system (Co thickness 1.6 nm and 5.8 nm) as coming from metallic Fe atoms at the interface. These authors also reported the 9% contribution of metallic Fe atoms in the XAS spectra. This is not the case in our XAS spectra, in which deposition of cobalt results only in a minor change in the line intensity at 708 eV, similar to $Fe_3O_4$, as shown in Fig. 5c and d. Therefore, we postulate that iron atoms at the interface remain in an oxidic state, presumably in the $Fe^{2+}$ form corresponding to this energy. Also, the XMCD spectrum, similar to the metallic Fe one, can be ascribed to FeO-like species [31,35,51]. Additionally, a CEMS spectrum measured for a sample with Co showed no trace of metallic Fe, whose contribution to the spectrum would be very distinct from hematite and detectable for one Fe monolayer.

## 4. Summary and Conclusions

We studied epitaxial heterostructures of ultrathin Co deposits on 10 nm magnetite $Fe_3O_4$(111) and hematite α-$Fe_2O_3$(0001) films. Using in situ STM, we showed that nucleation and growth morphology of cobalt is determined by the biphase superstructure artificially formed on the magnetite and naturally occurring on the hematite surfaces prepared in UHV. Initially, the Co deposit formed isolated nanoparticles that gradually coalesced above a nominal thickness of 0.4 nm. Finally, at the maximum studied thickness of 3 nm, a quasi-continuous film, with surface nano-islands three to four atomic layers high, was formed.

The XAS measurements showed that in both metal-oxide interfaces, approximately one monolayer of cobalt became oxidized. The XAS spectra at the $L_3$ Co edge indicated the formation of CoO; however, in the case of $Fe_3O_4$(111), the formation of interfacial cobalt spinel ($CoFe_2O_4$) cannot be excluded. The amount of oxidized cobalt only slightly exceeded 1/2 ML for the $Fe_3O_4$/Co interface and approached 1 ML for $\alpha$-$Fe_2O_3$. The XAS spectra at the $L_3$ Fe edge indicated that the oxidation of cobalt is accompanied by the interfacial reduction of iron. However, based on our CEMS measurements, we can exclude any bulk-like forms of Fe in a lower oxidation state, particularly metallic iron.

Using XMCD-PEEM, it was possible to image the magnetic domain structure with sub-micrometer resolution and elemental sensitivity. The magnetic domain structure for Co on magnetite is very fine and replicates the magnetic domain structure of the underlying FiM oxide. It is worth recalling at this point that for the given (111)-orientation of the magnetite films, the easy axes lie along four directions, none of which are in the film plane. This additional factor, alongside anti-phase domain boundaries, may contribute to the small size of the domains. Moreover, as indicated by CEMS measurements, all domains have a magnetization component perpendicular to the film. Such a domain structure can produce contrast in MFM measurements, which explains the results presented by Lewandowski et al. [39]. The identity of the domain structures of magnetite and cobalt suggests that the magnetization of cobalt also has an out-of-plane component. Cobalt and magnetite layers are ferromagnetically coupled, and an interfacial CoO layer does not hinder this coupling. From XMCD measurement as a function of the Co thickness, it is clear that the interfacial CoO layer is in the paramagnetic (or AFM) state at RT. In contrast, Co on hematite exhibits much larger, several micrometer-sized domains. Possibly, in this case, it is rather the FM domain structure of Co that determines the AFM domain structure of hematite. The observation of the XMCD contrast at the L-Fe edge in the nominally antiferromagnetic hematite film indicates the occurrence of uncompensated magnetic moments in the interface layer due to the proximity of cobalt.

The onset of the long range magnetic order is found at 0.5 nm Co on hematite and 0.2 nm Co on magnetite. This difference is explained by the larger extent of Co oxidation at the cobalt-hematite interface on one hand and by superparamagnetism stabilization at the cobalt-magnetite interface on the other hand.

Finally, from the XMCD spectra at the Co L-edge, we could determine the spin and orbital moment of Co as a function of the deposit thickness. Magnetic spin and orbital moments for our thickest deposit of 3 nm, $m_{spin}$ = 1.6 µB, $m_{orb}$ = 0.2 $\mu_B$ and $m_{ratio}$ = $m_{orb}/m_{spin}$ = 0.125 do not essentially deviate from bulk values. However, with decreasing thickness, an increase of magnetic moments is observed, especially significant of the orbital one, masked for Co on hematite by superparamagnetism. The limiting values of $m_{orb}$ = 0.47 $\mu_B$ and $m_{ratio}$ = 0.35 are significantly larger than the experimental and theoretical values in low-dimensional Co-metal systems.

We believe the present study significantly contributes to understanding the processes at the magnetic metal-oxide interfaces and paves the way for shaping their functional properties. Significantly, the Co/α-Fe$_2$O$_3$(0001)/Pt(111)/MgO(111) heterostructures are promising for driving the AFM spins in hematite, as recently suggested [24].

**Corresponding Author**

*Ewa Madej, ewa.madej@ikifp.edu.pl

**Authorship contribution statement - CRediT**

Conceptualization: EMa, JK, NS; Funding acquisition: JK, NS; Investigation: EMa, NK-M, KF, DW-Ś, MZ, NS; Methodology: EMa, NK-M, KF, JK, EMł, DW-Ś, MZ, JZ, NS; Software: DW-Ś, JZ; Supervision JK, NS; Validation: EMa, NK-M, KF, DW-Ś, MZ, NS; Visualization: EMa, NK-M, KF, JK, DW-Ś, JZ; Writing – original draft: EMa, JK; Writing – review and editing: EMa, NK-M, KF, JK, EMł, DW-Ś, MZ, JZ, NS;

**Acknowledgments**

This research was supported by the National Science Centre, Poland (NCN), grant number 2020/39/B/ST5/01838 and partially supported by the statutory research funds of ICSC PAS within the subsidy of the Ministry of Science and Higher Education, Poland.

# Supplementary material

# Interfacial chemistry meets magnetism: comparison of Co/Fe₃O₄ and Co/α-Fe₂O₃ epitaxial heterostructures.

*Ewa Madej* [*, 1], *Natalia Kwiatek-Maroszek* [1,2], *Kinga Freindl* [1], *Józef Korecki* [1], *Ewa Młyńczak* [1], *Dorota Wilgocka-Ślęzak* [1], *Marcin Zając* [2], *Jan Zawała* [1], *Nika Spiridis* [1]

[1] Jerzy Haber Institute of Catalysis and Surface Chemistry Polish Academy of Sciences, Niezapominajek 8, 30-239 Krakow, Poland

[2] National Synchrotron Radiation Centre SOLARIS, Jagiellonian University, Czerwone Maki 98, 30-392 Krakow, Poland

1. **CEMS analysis of the Co/hematite interface**.

CEMS analysis was performed ex-situ after synchrotron experiments. By shadowing different sample areas, it was possible to measure spectra for the hematite film without and with Co overlayer.

Fig. S1a shows a CEMS spectrum for the 10 nm hematite film without Co covered by 3 nm of MgO. The spectral component (96% of total spectral intensity) shadowed in magenta represents regular α-Fe₂O₃ sites with almost bulk hyperfine parameters. The purple component (4%) characterised by small magnetic hyperfine splitting comes from the interfaces with Pt and MgO, were Fe atoms meets nonmagnetic neighbours.

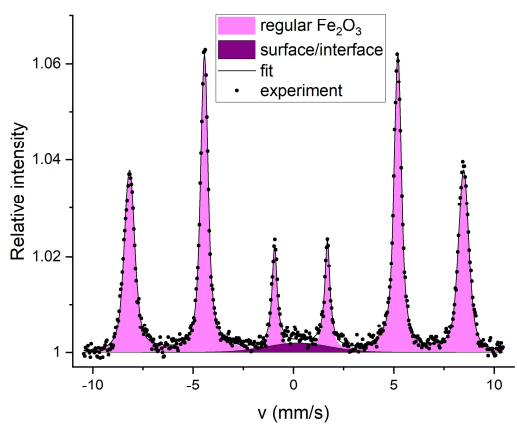

Fig. S1a CEMS spectrum for the 10 nm hematite film (without Co) covered by 3 nm of MgO.

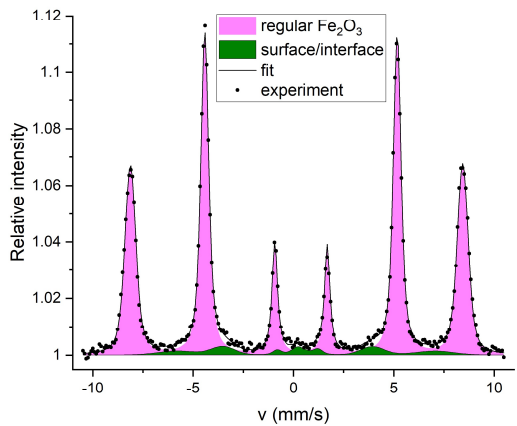

Fig. S1b CEMS spectrum for the 10 nm hematite film with 3 nm Co covered by 3 nm of MgO.

Figure S1b shows the spectrum from hematite covered with 3 nm of Co. The interfacial component is now more intense (7 %) and is characterised by a magnetic hyperfine field of 35 T and isomer shift of 0.51 mm/s. Such hyperfine parameters indicate Fe atoms in a lower oxidation state (most probably $Fe^{2+}$) and exclude the presence of metallic Fe. Metallic Fe component (5 % of spectral intensity) was simulated in Fig. S1c, which shows that this component would be distinctly seen at a velocity of 3 mm/s.

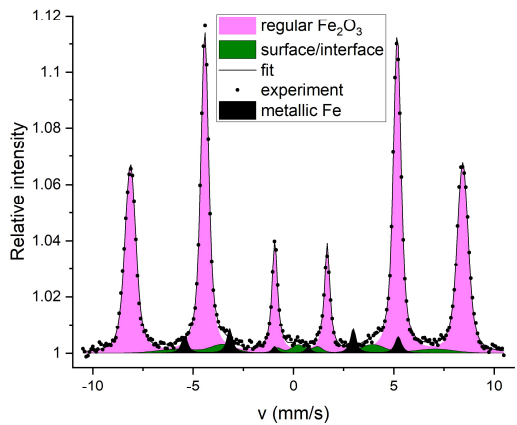

Fig. S1c CEMS spectrum for the 10 nm hematite film with 3 nm Co and simulated metallic Fe component.

2. **Magnetic domain structure**

Fig. S2a shows a XMCD-PEEM image measured at the Fe $L_3$ edge for the 10 nm magnetite film on the Pt buffer layer on MgO(111), covered with 3 nm of MgO. Magnetite domains have sub-micrometer sizes and irregular shapes.

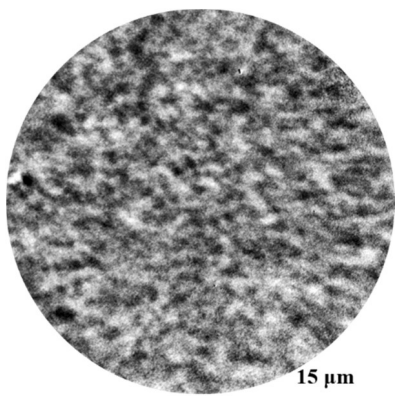

Fig. S2a Magnetic domains for 10 nm Fe$_3$O$_4$/Pt(111)/MgO(111), field of view 15 μm.

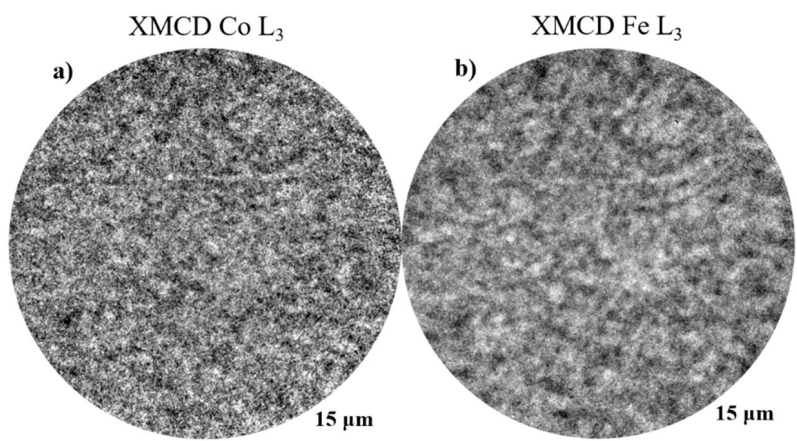

Fig. S2b XMCD-PEEM images (FoV 15 μm) recorded for 0.2 nm Co on 10 nm Fe$_3$O$_4$/Pt(111)/MgO(111) at the Co L$_3$ edge (a) and Fe L$_3$ edge (b).

3. **Kerr magnetometry**

The Kerr rotation as a function of the external magnetic field for 10 nm Fe$_3$O$_4$ film and 3 nm Co on magnetite is shown in Fig. S3a. Figure S3b shows magnetization curves for different cobalt thicknesses deposited on 10 nm α-Fe$_2$O$_3$ film on Pt(111)/MgO(111).

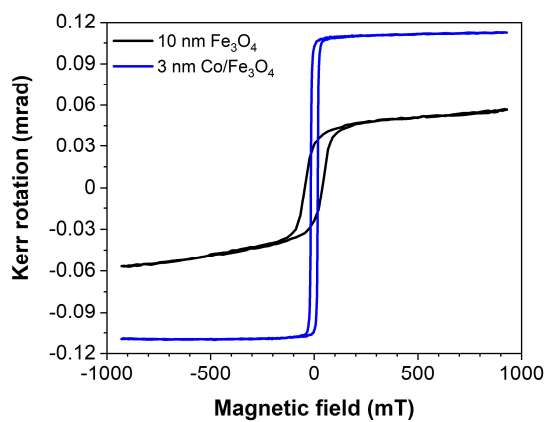

Fig. S3a Magnetization curves for 10 nm Fe$_3$O$_4$ film on Pt(111)/MgO(111) (black) and for 3 nm Co/Fe$_3$O$_4$/Pt(111)/MgO(111) (blue) obtained for p-polarized light.

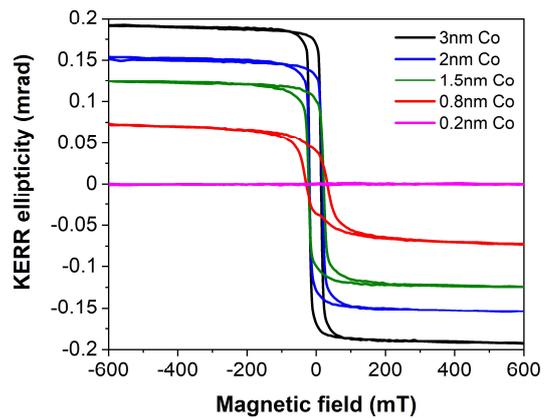

Fig. S3b Magnetization curves for different cobalt thicknesses deposited on 10 nm $Fe_2O_3$ film on Pt(111)/MgO(111) obtained for s-polarized light.